\documentstyle[12pt]{article}
\title{Energy-momentum tensor of quasiparticles in the effective gravity in
superfluids.}
\author{G.E. Volovik\\
Low Temperature Laboratory,
Helsinki University of Technology\\
P.O.Box 2200, FIN-02015 HUT, Finland\\
and\\
L.D. Landau Institute for Theoretical Physics,
 Moscow\\
}
\begin{document}
\maketitle
\begin{abstract}
{The problem of the energy-momentum conservation for matter in the
gravitational field is discussed on the example of the effective gravity,
which arises in superfluids. The "gravitational" field experienced by the
relativistic-like massless quasiparticles which form the "matter" (phonons in
superfluid 4He and low-energy fermions in superfluid 3He-A), is induced by
the flow of the superfluid "vacuum". It appears that
the energy-momentum conservation law for quasiparticles, has the covariant
form $T^{\mu}_{\nu;\mu}=0$. "Pseudotensor" of the energy-momentum for the
"gravitational  field" (superfluid condensate) appears to depend on "matter".
In the presence of the stationary "gravitational" (superfluid) field the real
thermodynamic temperature $T$ is constant in the true
dissipationless equilibrium state with no entropy production, while the
"relativistic" temperature
$T/\sqrt{g_{00}}$ is space dependent in agreement
with Tolman's law. In the presence of the event horizon
the true
dissipationless equilibrium state does not exist. The quasiequilibrium
dissipative motion across the horizon is considered.   The inflationary stage
of the expansion of the Universe can be modelled  using the expanding
Bose-condendsate.
 }
\end{abstract}

\tableofcontents

\section{\it  Introduction}  The effective gravity of the Sakharov type
\cite{Sakharov} can occur in many condensed matter systems. This
allows us to use these systems for modelling the specific properties of the
general relativity
\cite{UnruhSonic,Visser1997,Katanaev,Baush,JacobsonVolovik,Volovik1998,JacobsonV
olovik2}.
The important class of the efective gravity theories, which naturally arises
in condensed matter, is that which is induced by the gapless (massless)
quasiparticles. An example is the effective gravity in superfluid $^3$He-A
\cite{Volovik1998}. The energy spectrum of  fermionic
quasiparticles  in this superfluid  contains the topologically
stable point nodes (Fermi points)
\cite{Exotic}. In the vicinity of the gap node the quasiparticle spectrum
becomes fully relativistic
\begin{equation}
 g^{\mu\nu}(p_\mu -eA_\mu) (p_\nu -eA_\nu)=0~.
\label{rel}
\end{equation}
The topological stability of the gap nodes is crucial: the small
deformations of the quantum vacuum do not destroy the Fermi points but
deform their characteristics, the  slopes $g^{\mu\nu}$ and shifts
$A_\mu$  in Eq.(\ref{rel}), which play the part of the gravitational field and
different types of the gauge fields  correspondingly.  The  gravity and gauge
fields are thus the natural dynamical collective modes of the fermionic
quantum vacuum with Fermi points.

Using such an effective gravity we discuss the particular problems related to
the energy-momentum tensor of matter in the gravitational field.

\section{\it Hydrodynamics of superfluid liquid}

\subsection{\it  Basic
equations.}

  For the energy-momentum problem the
quasiparticle energy spectrum  only is important, so that the quantum
statistics of quasiparticles is irrelevant.  So, we can consider the
nomnrelativistic Landau two-fluid model of superfluidity, in which one
component of the fluid is formed by free quasiparticles  (Bose or Fermi)
moving on the background of the superfluid vacuum play the part of matter.
The effective gravitational field acting on quasiparticles is produced by the
motion of the another component of the liquid  -- the superfluid condensate,
which is characterized by the density
$\rho$ and superfluid velocity ${\bf v}_{(s)}$. The latter  is not necessarily
curl-free.  Let us recall some basic equations of two-fluid dynamics, usung
the simplest possible model with the following energy  of the liquid
\cite{Khalatnikov}:
\begin{equation}
{\cal E}=\int d^3r \left( {1\over 2}\rho{\bf v}_{(s)}^2 + \epsilon(\rho) +
\sum_{\bf p} E({\bf p},{\bf r}) f({\bf p},{\bf r})\right)~.
\label{Energy}
\end{equation}
Here the first term is the kinetic energy of the superfluid condensate;
the second term is its energy density as a function of $\rho$; $f({\bf
p},{\bf r})$ is the distribution function of quasiparticles; $\sum_{\bf
p}=\int d^3p/(2\pi\hbar)^3$ times the number of the spin degrees of freedom;
the quasiparticle energy spectrum is:
\begin{equation}
E({\bf p},{\bf r}) =E_{(0)}({\bf p},\rho({\bf r})) +  {\bf p}\cdot{\bf
v}_{(s)}({\bf r})~.
\label{QuasiparticleEnergy}
\end{equation}
where the second term is the Doppler shift in the moving condensate, and we
assumed that the quasiparticle energy
$E_{(0)}({\bf p},{\bf r})$  in the superfluid frame (in the frame where the
superfluid vacuum is at rest) depends on the coordinate only through the
density $\rho$.

The mass current density or the linear momentum density consists of the
vacuum current and the momentum of quasiparticles
\begin{equation}
 {\bf j}= \rho {\bf v}_{(s)} + {\bf P}~,~{\bf P}=\sum_{\bf p}
 {\bf p} f({\bf p},{\bf r})  ~ ,
\label{MassCurrentDensity}
\end{equation}
so that the continuity equation is
\begin{equation}
\dot \rho  + \vec\nabla (\rho {\bf v}_{(s)} + {\bf P}) =0~.
\label{ContinuityEquation}
\end{equation}
The superfluid velocity of the vacuum obeys the London equation:
\begin{equation}
\dot {\bf v}_{(s)} +  \vec\nabla {\delta {\cal E}\over \delta \rho}  -{{\bf
j}\over \rho} \times  (\vec\nabla  \times {\bf v}_{(s)})=0~.
\label{LondonEquation}
\end{equation}
In superfluid $^4$He the superfluid velocity is potential, so that vorticity
$\vec\nabla
\times {\bf v}_{(s)}$ is nonzero only in the presence of quantized vortices.
In superfluid
$^3$He-A the vorticity can be continuous and the London equation in this
form can be obtained if one neglects the axial anomaly and the dependence of
the energy on the anisotropy vector (see Eqs.(6.6)-(6.7) in Ref.
\cite{Exotic}). The last term in equation (\ref{LondonEquation}) states that
in the absence of anomalies the vorticity moves with the center-of-mass
velocity ${\bf j}/\rho$.

The distribution function $f$ of the
quasiparticles is  determined  by the kinetic equation:
\begin{equation}
\dot f - {\partial E\over \partial {\bf r}} \cdot {\partial f\over \partial
{\bf p}}+ {\partial E\over \partial {\bf p}} \cdot {\partial f\over \partial
{\bf r}}=J_{coll}~.
\label{KineticEq}
\end{equation}
The collision integral conserves the momentum and the energy   of
quasiparticles,  i.e.
\begin{equation}
\sum_{\bf p}  {\bf p}  J_{coll}=\sum_{\bf
p}  E_{(0)}({\bf p})  J_{coll}=\sum_{\bf
p}  E({\bf p})  J_{coll}=0~,
\label{ConservationCollision}
\end{equation}
but not
necessarily the particle number: as a rule the quasiparticle number is not
conserved in superfluids.

\subsection{\it  Momentum conservation}
From above equations one obtains the time evolution of the momentum density
for each of two subsystems: the superfluid background (vacuum) and
quasiparticles (matter). The momentum evolution of the superfluid vacuum is
\begin{equation}
\partial_t (\rho {\bf v}_{(s)})  =- \nabla_i(j_i{\bf v}_{(s)}) -\rho
\vec\nabla\left( {\partial \epsilon\over \partial \rho}  +\sum_{\bf p} f
{\partial E_{(0)}\over \partial \rho} \right) + P_i \vec\nabla  v_{(s)i}~.
\label{SuperfluidMomentumEq}
\end{equation}

Using the
kinetic equation Eq.(\ref{KineticEq}) and condition $\sum_{\bf p}  {\bf p}
J_{coll}=0$ from Eq.(\ref{ConservationCollision}), one obtains the evolution
of the  momentum density of quasiparticles:
\begin{equation}
\partial_t  {\bf P}   =\sum_{\bf p} {\bf p}  \partial_t f   =  -
\nabla_i(v_{(s)i}{\bf P})    -\nabla_i \left(  \sum_{\bf p} {\bf p}f
{\partial E_{(0)}\over
\partial p_i}\right)  -\sum_{\bf p}f
\vec\nabla E_{(0)} - P_i
\vec\nabla  v_{(s)i}~.
\label{QuasiparticleMomentumEq}
\end{equation}

Though the momentum of each subsystem is not conserved because of the
interaction with the other subsystem, the total momentum  is conserved:
\begin{equation}
\partial_t  j_i   = \partial_t (\rho   v_{(s)i} +P_i)=
-\nabla_i P_{ik}~,
\label{TotalMomentumEq}
\end{equation}
with the stress tensor
\begin{equation}
P_{ik} =  j_i v_{(s)k} + v_{(s)i} P_k+  \sum_{\bf p} p_k f {\partial
E_{(0)}\over
\partial p_i}  + \delta_{ik}G  ~,~G=\rho\left({\partial \epsilon\over \partial
\rho}+ \sum_{\bf p} f{\partial E_{(0)}\over \partial \rho}\right) -
\epsilon  ~.
\label{StressTensor}
\end{equation}

\section{\it  Quasiparticles in effective gravity field }

\subsection{\it  "Relativistic" quasiparticles }  Let us consider the
"relativistic" quasiparticles  -- phonons in superfluid
$^4$He or  fermionic quasiparticles in superfluid $^3$He. Their energy
spectrum is linear in the limit of low energy:
\begin{equation}
E({\bf p},{\bf r}) =cp +  {\bf p}\cdot{\bf v}_{(s)}~,~{\rm or}~(E-
{\bf p}\cdot{\bf v}_{(s)})^2-c^2p^2=0~.
\label{RelativisticQuasiparticleEnergy}
\end{equation}
This equation is correct for phonons in superfluid
$^4$He, with $c$ being the speed of sound. In the $^3$He-A the
low energy spectrum is also linear but  anisotropic (see \cite{Exotic}). The
simplified equation (\ref{RelativisticQuasiparticleEnergy}) is valid in
$^3$He-A only for the homogeneous order parameter and is obtained after
shifting of the momentum and rescaling. The function $c$ in this case is
some combination of the slopes of the energy spectrum.

\subsection{\it  Effective metric}
Since the energy spectrum has the
relativistic form, let us rewrite the time  evolution of the quasiparticle
momentum density in Eq.(\ref{QuasiparticleMomentumEq}) in the covariant form,
using the superfluid system as the effective background metric. The
corresponding metric follows from the quasiparticle spectrum  in
Eq.(\ref{RelativisticQuasiparticleEnergy}), which can be written in a
general Lorentzian form
\begin{equation}
g^{\mu\nu}p_\mu p_\nu=0~,
\label{RelativisticSpectrum}
\end{equation}
where the metric elements are
\begin{equation}
g^{00}=1~,  ~g^{0i}=v_{s}^i~,~
g^{ik}=- (c^2\delta^{ik}-v_{(s)}^iv_{(s)}^k)~.
\label{gikUp}
\end{equation}
and $p_0=-E$, $p^0=-E_{(0)}=-cp$, $p^\alpha p_\alpha=  E_{(0)}^2- c^2p^2$.

The  metric of the effective
space in which  quasiparticles  propagate along the geodesics, i.e.
$ds^2=g_{\mu\nu}dx^\mu dx^\nu=0$, is correspondingly
\begin{equation}
g_{00}=\left(1 - {v_{(s)}^2\over c^2}\right)~,  ~g_{0i}={v_{(s)i}\over
c^2}~,~ g_{ik}=- {1\over c^2}\delta_{ik} ~.
\label{gikDown}
\end{equation}
Except for the conformal factor, it coincides with acoustic metric discussed
 for sound propagation in normal \cite{UnruhSonic,Visser1997}
and superfluid  \cite{Ilinski} liquids. In our case the
quasiparticles are not necessarily the phonons.
The kinetic equation acquires the relativistic form
\begin{equation}
p^\alpha {\partial f\over
\partial x^\alpha}-\Gamma^\alpha_{\beta\gamma}p^\beta p^\gamma  {\partial
f\over
\partial p^\alpha}=\tilde J_{coll}~,~\tilde J_{coll}=p^0
J_{coll}~,~p^0 = -E_{(0)}~.
\label{KineticEqRel}
\end{equation}

The system of quasiparticles play the part of the matter in general
relativity. That is why one can expect that the momentum "conservation law"
for matter has the usual covariant form
\begin{equation}
 T^{\mu}_{\nu;\mu}=0~,~ ~{\rm or}~~ {1\over
\sqrt{-g}}\partial_\mu \left(T^{\mu}_\nu \sqrt{-g}\right) - {1\over
2}T^{\alpha\beta} \partial_\nu g_{\alpha\beta}= 0~,
\label{Nonconservation1}
\end{equation}
where $\sqrt{-g}=c^{-3}$. The terms which are not the total derivatives
represent the forces acting on the matter from the effective gravitational
field. Let us check  if this equation can be applied to our case, where the
effective gravity is induced by the superflow.

\subsection{\it Energy-momentum tensor}
Let us introduce the components of the energy-momentum tensor:
\begin{eqnarray}
\sqrt{-g} T^{0}_{i}=-P_i ~~,~\sqrt{-g} T^{0}_0=\sum_{\bf p}   f E
~,~\sqrt{-g} T_{i}^k= - \sum_{\bf p} p_i f {\partial E\over \partial p_k}
~,\nonumber
\\ \sqrt{-g}T_{0}^i= P^ic^2 +v_{(s)}^i\sum_{\bf p}   f E_{(0)}+ v_{(s)}^k
\sum_{\bf p} p_k f {\partial E\over \partial p_i} ~.
\label{Tupdown}
\end{eqnarray}
\begin{eqnarray}
  \sqrt{-g}T^{00}= \sum_{\bf p}   f E_{(0)}
~,~ \sqrt{-g}T^{0i}=c^2P^i + v_{(s)}^i\sum_{\bf p}   f E_{(0)}
~,\nonumber
\\ \sqrt{-g} T^{ik} = c^2 \sum_{\bf p} p_i f {\partial E_{(0)}\over
\partial p_k} + c^2(v_{(s)}^kP^i +v_{(s)}^iP^k) +v_{(s)}^i v_{(s)}^k
\sum_{\bf p}   f E_{(0)}~.
\label{Tupup}
\end{eqnarray}

 Using the
above expression for the quasiparticle energy-momentum tensor,   one can be
rewrite  the equation for the quasiparticle momentum density,
Eq.(\ref{QuasiparticleMomentumEq}), and the analogous
equation for the  evolution of quasiparticles energy density in the following
form:
\begin{equation}
  \partial_\mu \left( \sqrt{-g} T^{\mu}_\nu\right) - {\bf P}\cdot
\partial_\nu {\bf v}_{(s)} -
 { \partial_\nu c \over c}\sum_{\bf p}   f E_{(0)}  = 0~.
\label{ForceFromSuperflow}
\end{equation}
The last two terms are the forces acting on the subsystem of quasiparticles
from the superfluid background, if the latter is inhomogeneous. These forces
exactly reproduce the forces acting on the matter from the gravitational
field in Eq.(\ref{Nonconservation1}):
\begin{equation}
{1\over 2}\sqrt{-g}
T^{\alpha\beta} \partial_\nu g_{\alpha\beta}= {\bf P}\cdot
\partial_\nu {\bf v}_{(s)} +
 { \partial_\nu c \over c}\sum_{\bf p}   f E_{(0)}  ~.
\label{Forces}
\end{equation}
Thus the energy-momentum tensor of quasiparticles satisfies the covariant
equation (\ref{Nonconservation1}). This result does not depend on the
dynamic properties of the superfluid condensate (gravity field) and follows
only from the "relativistic" spectrum of quasiparticles.

\section{\it Thermal equilibrium}

\subsection{\it True temperature and "relativistic" temperature}

In thermal equilibrium the quasiparticle
distribution function is
\begin{equation}
f=  {1\over \exp {E({\bf p},{\bf r}) - {\bf p}\cdot{\bf
v}_{(n)}\over T} \pm 1  }~.
\label{ThermalEquilibrium}
\end{equation}
 It is determined by the temperature $T$ of the liquid and by the
velocity ${\bf v}_{(n)}$ of the normal component of the liquid represented
by the quasiparticles.  This form of the distribution function is determined
by the conservation of the energy and momentum by the
collision term in kinetic equation. The Doppler shifted energy is
\begin{equation}
E({\bf p},{\bf r}) - {\bf p}\cdot{\bf v}_{(n)}=cp -  {\bf p}\cdot{\bf
w} ~,~{\bf w}={\bf v}_{(n)} - {\bf v}_{(s)}
\label{Counterflow}
\end{equation}
${\bf w}$ is called the velocity of counterflow  of
the normal and superfluid fractions.
Using these two velocities one can introduce the 4-velocities of the
"matter",  $u^\alpha$ and
$u_\alpha=g_{\alpha\beta}u^{\beta}$, which satisfy equation
$u_\alpha u^\alpha=1$:
\begin{equation}
u^0={1\over \sqrt{ 1 - {w^2\over
c^2}}}~,~u^i={v_{(n)}^i\over \sqrt{ 1 - {w^2\over
c^2}}}~~,~u_i= -{ w_i \over c^2\sqrt{ 1 -
{w^2\over c^2}}}~,~u_0={1+{{\bf w} \cdot{\bf v}_{(s)}\over c^2} \over
\sqrt{ 1 - {w^2\over c^2}}}~,
\label{4Velocity}
\end{equation}
In terms of the 4-velocity the equilibrium distribution function has the
relativistic covariant form
\begin{equation}
f=  {1\over \exp (-p_\mu u^\mu / T_{\rm eff}) \pm 1  }~,~  T_{\rm
eff}={T\over \sqrt{ 1 - {w^2\over c^2}}}~.
\label{ThermalEquilibriumRel}
\end{equation}
The temperature
$T$ is a real thermodynamic temperature, while the covariant "relativistic"
temperature $ T_{\rm eff}$ is not. This follows from the conditions for the
global equilibrium, which means the absence of dissipation and entropy
production.

\subsection{\it Global equilibrium in nonrelativistic liquid and in
relativistic thermodynamics.}

The global equilibrium state in
nonrelativistic superfluids is determined by the following  conditions for the
absence of dissipation \cite{Khalatnikov,UFN}:
\begin{eqnarray}
T=const ~,
\label{EquilibriumConditionT}\\
{\bf v}_{(n)}={\bf a} + {\bf b}\times
{\bf r}~,~{\bf a} =const ~,~  {\bf b}=const ~,
\label{EquilibriumConditionVn}\\
\vec\nabla\cdot ({\bf j} -\rho {\bf v}_{(n)})=0~,
\label{EquilibriumConditionJ}\\
{\delta {\cal E}\over \delta \rho}+{\bf v}_{(s)}\cdot
{\bf v}_{(n)}=const ~.
\label{EquilibriumConditionMu}
\end{eqnarray}

In the presence of the
stationary inhomogeneous "gravitational field" (i.e. in the presence of the
fields
${\bf v}_{(s)}={\bf v}_{(s)}({\bf r})$,  $c=c({\bf r})$ and  $\rho=\rho({\bf
r})$) the last 3 conditions require
${\bf v}_{(n)}=0$ in the frame where the superfluid (gravity) fields do not
depend on time.
Let us consider this on  example of Eq.(\ref{EquilibriumConditionJ}).
According to continuity equation Eq.(\ref{ContinuityEquation}), in the
stationary case one has
$\vec\nabla\cdot{\bf j}=0$. Then using the
Eq.(\ref{EquilibriumConditionVn}), one can see that the  condition in
Eq.(\ref{EquilibriumConditionJ}) becomes
$ {\bf v}_{(n)}\cdot\vec\nabla  \rho =0$. Thus if  $\rho$ depends on ${\bf
r}$,  the requirement for equilibrium is ${\bf v}_{(n)}=0$.
Then if $ {\bf v}_{(s)}$ depends on ${\bf
r}$, the condition ${\bf v}_{(n)}=0$ follows from
Eq.(\ref{EquilibriumConditionMu}). Thus in the "gravitational field"  (${\bf
v}_{(s)}={\bf v}_{(s)}({\bf r})$,   $\rho=\rho({\bf r})$) one always has
${\bf v}_{(n)}=0$ as the global equilibrium condition. In general it means
that if any order parameter texture moves with respect to the normal
component, there is a dissipative friction between the texture and
quasiparticle system.

Thus the true equilibrium (nondissipative) state in the presence of the
stationary  "gravitational field" is characterized by the constant real
temperature $T$, while the covariant "relativistic" temperature
changes in space as $ T_{\rm eff}({\bf r})=T/\sqrt{ 1 -  {\bf v}_{(s)}^2({\bf
r})/ c^2({\bf r})}= T/  \sqrt{ g_{00}({\bf r})}$ in agreement
with Tolman's law \cite{Tolman}. This will be illustrated further below for
the one-dimensional texture.

Without the "gravity" field, i.e. at constant $\rho$, $c$ and ${\bf
v}_{(s)}$, the 6 equilibrium conditions above can be expressed in the
relativistic form using the  relativistic temperature vector
$\beta_\mu=u_\mu/ T_{\rm eff}$:
\begin{equation}
\partial_\nu \beta_\mu +\partial_\mu \beta_\nu =0~,~\beta_\mu={u_\mu\over
T_{\rm eff}}~.
\label{TemperatureVector1}
\end{equation}
The covariant extension of the global
equilibrium condition in Eq.(\ref{TemperatureVector1}) to the case of the
stationary "gravity" field (i.e. $\rho$, $c$ and ${\bf v}_{(s)}$ are
functions of ${\bf r}$) is that the temperature vector $\beta_\mu$ is the
timelike Killing-vector (see Ref.\cite{Zimdahl} and references there):
\begin{equation}
 \beta_{\mu;\nu}+ \beta_{\nu;\mu}=0~.
\label{TemperatureVector2}
\end{equation}
It is easy to check for the 1-dimensional case discussed later, that
the condition  ${\bf v}_{(n)}=0$ and $T=const$ for the global equilibrium of
the liquid is   consistent with the covariant global equilibrium
equation (\ref{TemperatureVector2}), unless an event horizon is present.
The thermodynamic equilibrium in the presence of the black hole horizon
has been discussed in detail in \cite{Hayward}.

\subsection{\it Energy-momentum tensor in quasiequilibrium.}

In quasiequilibrium states $T$ and ${\bf v}_{(n)}$ are the local hydrodynamic
quantities which depend on space and time coordinates.
The quasiparticle momentum density in a local equilibrium is
\begin{equation}
{\bf P}({\bf r},t)=\sum_{\bf p}
 {\bf p} f({\bf p},{\bf r},t)=\rho_n^{(0)}{ {\bf w}\over \left(1 - {w^2\over
c^2}\right)^3} ~ ,
\label{EquilMassCurrentDensity}
\end{equation}
where $\rho_n^{(0)}$ is normal fluid density in the limit of zero
counterflow, ${\bf w}=0$. For the spin-$1/2$ fermions
\begin{equation}
\rho_n^{(0)}={7\over 90}\pi^2 {T^4\over c^5}= {7\over 90}\pi^2 {T^4\over
c^2}  \sqrt{-g} ~ .
\label{NormalDensity}
\end{equation}
The quasiparticle energy density is
\begin{equation}
 \sum_{\bf p}
E_{(0)} f ={\varepsilon^{(q)} + P^{(q)}\over  1 - {w^2\over
c^2}}-P^{(q)}= {3\over 4} \rho_n^{(0)}c^2{  1+ {1\over 3}{w^2\over
c^2} \over \left(1 - {w^2\over
c^2}\right)^3} ~ ,
\label{EquilEnergytDensity}
\end{equation}
where $P^{(q)}$ is the quasiparticle contribution to pressure
\begin{equation}
P^{(q)} =   {1\over 4}{ \rho_n^{(0)}c^2\over \left(1 - {w^2\over
c^2}\right)^2}~,~{\partial P^{(q)}\over \partial {\bf v}_n}={\partial
P^{(q)}\over \partial {\bf w}}={\bf P}~.
\label{QuasipPressure}
\end{equation}
The quasiparticle contribution to the stress tensor is
\begin{equation}
 \sum_{\bf p} p_k f {\partial
E_{(0)}\over \partial p_i} = (\varepsilon^{(q)} + P^{(q)}){w_iw_k\over  c^2 -
w^2}+P^{(q)}\delta_{k}^i~.
\label{QuasipStressTensor}
\end{equation}
where  $ \varepsilon^{(q)} =3 P^{(q)}=\rho^{(q)}c^2$ is the energy density
of "matter" and $\rho^{(q)}$ is its "mass".

Combining the above equations, one can write the energy-momentum tensor for
quasiparticles in the general "relativistic" form using the
4-velocity introduced in Eq.(\ref{4Velocity}):
\begin{equation}
T^{\mu\nu}  =   (\varepsilon  + P )u^\mu
u^\nu-P g^{\mu\nu}~ .
\label{QuasipStressTensorRel2}
\end{equation}
where $\varepsilon$ is the "covarant" energy density of quasiparticles:
$\varepsilon =  \varepsilon^{(q)}/
\sqrt{-g}$. For the
$1/2$-spin fermionic quasipaticles one has
\begin{equation}
  \varepsilon = {\varepsilon^{(q)}\over
\sqrt{-g}}={7\over 120}\pi^2 T_{\rm eff}^4={7\over 120}\pi^2 {T^4 \over
\left(1 - {w^2\over c^2}\right)^2} ~,~ P ={\varepsilon\over 3}~.
\label{QuasiparticleEnergyDensity}
\end{equation}

The relevant components of the energy-momentum tensor are
\begin{eqnarray}
 T^{0}_{i}=  -(\varepsilon + P) {w_i\over c^2-w^2} ~,~ T_{i}^k=-(\varepsilon
+ P ) {v_{(n)}^k w_i\over  c^2-w^2}  - P \delta_i^k~, \nonumber\\
T^{i}_{0}=
(\varepsilon + P)  v_{(n)}^i {c^2+ {\bf w}\cdot{\bf v}_{(s)}\over c^2-w^2}~,~
T^{00} =
{(\varepsilon + P)  \over 1-w^2/c^2} -P~,~ T^{0}_{0}=T^{00} +  {{\bf P}\cdot
{\bf v}_{(s)}\over \sqrt{-g}} ~.
\label{Tequilibrium}
\end{eqnarray}

\subsection{\it The energy-momentum pseudotensor}

The equation for the energy-momentum of quasiparticles is covariant but this
equation does not represent any conservation law. On the contrary, the total
energy and momentum of the system,  quasiparticles + superfluid condensate
(gravity field), are conserved, but this conservation laws cannot be
expressed in the covariant form. This is another problem of the general
relativity, which probably also can be attacked using the condensed matter
analogy. In the condensed matter the energy-momentum tensor of the system
exists on the microscopic level, i.e. it can be expressed using the
microscopic variables. But in many cases, for example in ferromagnets
\cite{Volovik1987}, it cannot be expressed in terms of the local
effective variables, such as magnetization. The effective gravity field as
the  local low-energy collective mode can be just an another example.

Let us consider the analog of the pseudotensor of the energy-momentum of the
gravitational field, which is in our case the part of the total energy
momentum tensor coming from the superfluid motion of the condensate.
The conservation law for the total energy-momentum can be written as
\begin{equation}
  \partial_\mu \left( \sqrt{-g} T^{\mu}_\nu  +\sqrt{-g}
t^{\mu}_\nu\right)=0~,
\label{Conservation}
\end{equation}
where $t^{\mu}_\nu$ is  the pseudotensor of energy-momentum.
From the basic equations for superfluids it follows that
\begin{eqnarray}
 \sqrt{-g}t^{0}_{i}=  -\rho v_{(s)i} ~,  \nonumber \\
\sqrt{-g}t^{k}_{i}=
-j^k v_{(s)i} -\delta^k_i \left[\rho\left({\partial\epsilon\over
\partial\rho}   +
\sum_{\bf p} f  {\partial E_{(0)}\over \partial\rho}\right) -\epsilon\right]
~, \nonumber
\\
\sqrt{-g}t^{0}_{0}={1\over 2}\rho{\bf v}_{(s)}^2 + \epsilon(\rho)
~, \nonumber \\
 \sqrt{-g}t^{i}_{0}= j^i\left({\partial\epsilon\over
\partial\rho} +
\sum_{\bf p} f  {\partial E_{(0)}\over \partial\rho} + {{\bf
v}_{(s)}^2\over 2}\right)~.
\label{pseudotensor}
\end{eqnarray}
It appears, that the pseudotensor of the "gravitational field" depends  on
matter.

Let us introduce the Lagrangian
\begin{equation}
L= {1\over 2}\rho{\bf v}_{(s)}^2 + \epsilon(\rho) + \sum_{\bf
p} E  f +\rho\partial_t \Phi ~.
\label{ Lagrangian}
\end{equation}
where  $\Phi$ is the phase of the condensate related to the superfluid
velocity:
\begin{equation}
  \vec\nabla \Phi={m\over \hbar}{\bf v}_{(s)}~.
\label{Phase}
\end{equation}
Here $m$ is the mass of the initial boson in the superfluid Bose
condensate, and further we put $\hbar/m=1$. Then the above pseudotensor of
the "gravitational field" can be written in the compact form:
\begin{equation}
  \sqrt{-g}t^{\mu}_{\nu}=L\delta^{\mu}_{\nu} - {\partial L\over \partial
\nabla_\mu  \Phi} \nabla_\nu  \Phi -\delta^{\mu}_{\nu} \sum_{\bf
p} E  f  ~.
\label{pseudotensorGeneral}
\end{equation}

\section{\it Motion of normal component through horizon}

\subsection{\it 1-dimensional stationary "gravity" field}

Let the superfluid-gravity field is stationary and depends  only
on one space coordinate $x$.
The motion of the "matter" is determined by Eq.(\ref{ForceFromSuperflow}),
which using Eq.(\ref{Tequilibrium}) and ${\bf
w} =\hat{\bf x} w(x)$,  ${\bf
v}_{(s)}=\hat{\bf x}v(x)$ read:
\begin{eqnarray}
\sqrt{-g} T^{x}_{0}=const=4  P^{(q)} ~{(v+w)(c^2+wv)\over
c^2-w^2}~,
\label{horizon1}\\
-\partial_x \left( 4 P^{(q)}~{(v+w)w\over
c^2-w^2} + P^{(q)}\right )=  { 4 P^{(q)}w\over
c^2-w^2}\partial_x v +
 {\partial_x c\over c} ~\left( 4 P^{(q)}~{c^2\over
c^2-w^2} - P^{(q)}\right ).
\label{horizon2}
\end{eqnarray}
The Eq.(\ref{horizon1}) manifests the constant energy flux along $x$ axis.
The flux is zero if the normal component velocity is at rest in the frame of
the texture, i.e. if ${\bf v}_{(n)}=0$, or $w(x)=-v(x)$. In this case the
second equation Eq.(\ref{horizon2}) is satisfied if the temperature $T$ is
constant, so that
\begin{equation}
P^{(q)}(x)\propto
{T^4\over c(x)^3}\left(1-{v^2(x) \over c^2(x)}\right)^{-2} ~.
\label{TrueEquilibrium}
\end{equation}
 This solution describes  a true equilibrium
state in the presence of the "gravity field": there is no dissipation (the
conditions for the absence of dissipation is
${\bf v}_{(n)}=0$ and $T=const$). Note that the relativistic temperature
$T_{\rm eff}$ depends on $x$ in this equilibrium state. The
covariant equilibrium condition $\beta_{\mu;\nu}=0$ is satisfied.

\subsection{\it Absence of the global equilibrium state in the presence of
horizon}

Let us consider the motion of quasiparticle system in the presence of the
event horizon. For simplicity we assume the constant velocity ${\bf
v}_{(s)}=\hat{\bf x}v$ and the coordinate dependent speed of sound $c(x)$,
so that at some point $x=x_0$ one has $c^2(x_0)=v^2$. Since $g_{00}= 1
-  v_{(s)}^2/ c^2$
in Eq.(\ref{gikDown}) changes sign at $x=x_0$ one has a
horizon at this point. In the
simplest realization of the horizon the speed of "light" changes across the
horizon in the following way:
\begin{equation}
 c(x) =v(1-a \tanh (bx))~.
\label{c(x)AcrossHorizon}
\end{equation}
Here $a$ and $b$ are constant parameters and the position of horizon is
chosen at $x=0$.

Outside the horizon one can determine the
true equilibrium state with ${\bf
v}_{(n)}=0$ (or $w=-v$) and $T=const$. However, this state cannot be
determined globally is the whole space: the "relativistic"
temperature
\begin{equation}
 T_{\rm eff}= { T  \over \sqrt{ g_{00}}}~,~  g_{00} =1 - {v^2\over c^2}~,
\label{EffectiveT}
\end{equation}
 which determines the
"relativistic" pressure $P^{(q)}$, diverges at the horizon
together with the energy density and cannot be extended across the horizon.
Thus in the presence of a horizon, one must  look for the
quasiequilibrium solution with the space dependent $T(x)$ and
$w(x)$ and thus with dissipation.

In general relativity the vacuum energy (the energy of the Boulware vacuum)
also diverges at horizon, but with the negative sign. In the Hartle-Hawking
state, in which the thermodynamic temperature $T$ equals the Hawking
temperature
$T_{\rm H}=(\hbar/2\pi) (dc/dx)|_{\rm hor}$ the two divering terms in the
stress-energy tensor cancel each other (see references in the recent paper
\cite{Mukohyama}). However in condensed matter
such a diveregence of the energy density in each of the two subsystems
(quasiparticles, which form the "matter", and superfluid condensate, which
produces the effective  gravity field) represents the real physical
singularity at horizon. The external observer
who lives in the Galilean world of the physical laboratory  can use
the "superluminal" signals to measure the quasiparticle
distribution function both outside and inside the horizon. He will find that
the quasiparticle energy density does really diverge at horizon, so that the
global equilibrium does not exist. Such a singularity can be avoided either
by escaping from the global equilibrium to the local one with dissipation, or
by consideration of the high-energy nonrelativistic corrections to the
quasiparticle spectrum, together with the reconstruction of the superfluid
vacuum within the horizon, which again leads to dissipation.

\subsection{\it Dissipative motion of normal component across horizon.}

In the presence of horizon, the energy flux in Eq.(\ref{horizon1}) is no more
zero and one can express  $ P^{(q)}$ from this equation
\begin{equation}
P^{(q)}\propto {c^2-w^2\over (v+w)(c^2 +vw) } ~.
\label{P}
\end{equation}
and insert  it to  Eq.(\ref{horizon2}) to  obtain the following equation
for $w$ (note that we assumed $v=const$ for the superfluid velocity):
\begin{equation}
-\partial_x \left({4vw +3w^2 +c^2\over (v+w)(c^2 +vw) }\right)=
{\partial_x c\over c} ~\left({w^2 +3c^2\over (v+w)(c^2 +vw) }\right)~.
\label{w(x)}
\end{equation}
It follows that $w$ depend on $x$ as a function of $c(x)$, i.e.
$w=w(c(x))$, where $w(c)$ satisfies the equation
\begin{equation}
-c{d\over dc} \left({3w \over c^2 +vw  }+ {1\over v+w  }\right)=
 {w^2 +3c^2\over (v+w)(c^2 +vw) }~.
\label{w(c)}
\end{equation}

Introducing $c=vC$ and $w=W vC$,  one obtains
\begin{equation}
-{d\over dC}\left({3W\over  C+W  }+ {1\over 1 +CW  }\right)=
{W^2 +3 \over (1+CW)(C+W) } ~.
\label{W}
\end{equation}
This equation has a particular solution $W=-1$, or $w=-c$, which according
to Eq.(\ref{P}) corresponds to zero temperature, $T=0$.

The function $C(x)$, which gives the horizon at $x=0$ in
Eq.(\ref{c(x)AcrossHorizon}), is
\begin{equation}
 C =1 -  a \tanh (bx)~.
\label{C}
\end{equation}
If $a\ll 1$ and thus $|C-1|\ll 1$, one has the following solution of this
equation
\begin{equation}
 W = -{3 \over 2} (C-1)  ~.
\label{SolutionSmallJump}
\end{equation}
This solution has no singularity at the horizon ($x=0$). The
temperature $T$ and the counterflow velocity $w$ slightly change  across
the horizon.
Since the normal velocity $v_{(n)}=w+v_{(s)}\approx v_{(s)}$ is nonzero in the
horizon frame, this state in Eq.(\ref{SolutionSmallJump}) is in local thermal
equilibrium, but not in a global one. Therefore there is a dissipation.

Let us consider another extreme case $|W+1|\ll 1$, when the velocity
of the normal component is small in the horizon frame. This is close to
the Boulware vacuum in the general relativity. In this regime the solution of
Eq.(\ref{W}) is
\begin{equation}
 W = -1+A (C-1)^2  ~,
\label{GuessFunction}
\end{equation}
where $A$ is an arbitrary parameter and it is assumed that $|C-1|\ll 1$.
Since $c^2-w^2\propto A (c-v)^2$, $v+w  \propto - (c-v)$ and $c^2+wv\propto
(c-v)$, one has $T^4  \propto (c^2-w^2)^3/(v+w)(c^2+wv) \propto (c-v)^4$,
i.e.  $T  \propto |c-v|$ and  $ T_{\rm eff}=T/\sqrt{1-w^2/c^2}  =const$.
In this solution the real temperature $T$ is space dependent and is zero at
horizon, while the effective relativistic temperature $ T_{\rm eff}$ is
constant across the horizon.

\section{\it Simulation of Hubble expansion}

The expanding solution of the motion equations is
\begin{equation}
 {\bf v}_{(s)}({\bf r}, t)={\bf v}_{(n)}({\bf r}, t)=\tilde H(t){\bf
r}~,~\rho=\rho(t)~.
\label{Hubble expansion}
\end{equation}
The Hubble parameter and its connection to the density of the liquid are found
from Eqs.(\ref{ContinuityEquation},\ref{LondonEquation}):
\begin{equation}
 \tilde H(t)= -{\partial_t \rho \over 3\rho}~,~  \partial_t \tilde H
+\tilde H^2=0~.
\label{HubbleParameterEquation}
\end{equation}
which give
\begin{equation}
 \tilde H(t)=   (t-t_0)^{-1}~,~   \rho(t)= {\rm Const}~ (t-t_0)^{-3}~
\label{HubbleParameter}
\end{equation}
Note that the Hubble parameter in the radiation-dominated Universe is
$H(t)=(1/2)(t-t_0)^{-1}= (1/2)\tilde H(t)$.

Since the quasiparticle momentum ${\bf P}=0$ (because ${\bf v}_{(s)} ={\bf
v}_{(n)}$), the Eq.(\ref{QuasiparticleMomentumEq}) is automatically
satisfied, while the energy conservation law for quasiparticles gives the
following equation for the quasiparticles energy density
$ \varepsilon^{(q)}=3 P^{(q)}( =\rho^{(q)}c^2)$:
\begin{equation}
 \partial_t \varepsilon^{(q)}  +\left(4\tilde H-  {\partial_t c\over
c}\right)\varepsilon^{(q)}=0~.
\label{EnergyEquation}
\end{equation}

 Since $\varepsilon^{(q)}=\varepsilon/c^3 \propto
T^4/c^3$, one obtains equation for
$T$:
\begin{equation}
 \partial_t \left({T\over c}\right)  +\tilde H \left({T\over c}\right)=0~.
\label{TEquation}
\end{equation}
which gives $T(t)\propto c(t)/(t-t_0)$.

The time dependence of the speed of
light is determined by its dependence on the density $\rho$:
$c(t)=c(\rho(t))$.  The cosmological scenario of the expansion of the
radiation-dominated Universe, with $T(t)\propto (t-t_0)^{-1/2}$, is obtained
if
$c\propto (t-t_0)^{1/2}$. If the speed $c$ comes from the compressibility of
the superfluid vacuum, then such behavior of $c$ can result from the
following equation of states for the superfluid condensate:
$\epsilon(\rho) \sim
\rho^{2/3}$ which gives $c(\rho) \sim \rho^{-1/6}\propto (t-t_0)^{1/2}$.
Note that in the cases of the Bose condensation of the almost ideal Bose
gas and of the superfluidity of the almost ideal Fermi
gas, the equation of states are corespondingly
$\epsilon(\rho) \sim \rho^{2}$ and $\epsilon(\rho) \sim \rho^{5/3}$ with
$c(\rho) \sim \rho^{1/2}\propto (t-t_0)^{-3/2}$ and $c(\rho) \sim \rho^{1/3}
\propto (t-t_0)^{-1}$.

If one introduces the comoving coordinate frame: ${\bf r}=\tilde{\bf r}
\tilde a(t)$ with $\partial_t \tilde a=\tilde a\tilde H$, one obtains the
following effective metric in this frame
\begin{equation}
ds^2=dt^2 -a^2 d\tilde{\bf r}^2~,~a={\tilde a\over c}~,~{\partial_t \tilde
a\over
\tilde a}=\tilde H~.
\label{MetricComovingFrame}
\end{equation}
Since $\tilde a\propto t-t_0$, the behavior corresponding to   the
radiation-dominated Universe occurs if $c\propto
(t-t_0)^{1/2}$. Then
$a=\tilde a/c
\propto (t-t_0)^{1/2}$ and
$\partial_t a/a=H=(1/2)\tilde H$. Such case occurs if $\epsilon(\rho) \sim
\rho^{2/3}$.

Another special case is when  $c\propto  t-t_0$. In this case
$a={\rm Const}$, and $T={\rm Const}$. This means that the effective
space-time for quasiparticles is Minkowski flat space-time, inspite of the
nontrivial dynamics of the background superfluid condensate. Such case occurs
if $\epsilon(\rho) \sim \rho^{1/3}$ which gives $c(\rho) \sim
\rho^{-1/3}\propto t-t_0 $.

In the case of the superfluidity of the almost ideal Bose
and Fermi gases, the integral $\int dt ~a^{-1}(t)$ diverges at the origin.
Thus both cases correspond to the power-law inflation.

\section{\it Conclusion.}

The next step is to derive the effective action for the "gravity field". The
effective gravity theory is obtained by integration over the fermionic
degrees of freedom in the presence of the background field
\cite{Sakharov}. The latter field becomes dynamical and represents the
low-frequency collective modes of the system of interacting particles  --
$^3$He or $^4$He atoms. In the case discussed here the result is well known:
the hydrodynamic equations for the background superfluid vacuum in
Eqs.(\ref{ContinuityEquation},\ref{LondonEquation}), which do not depend
on the details of the microscopic interactions between the atoms, but
which are far from being covariant and relativistic. To have the relativistic
equations for the effective gravity field one must consider such
hypothetical condensed mater systems where the main contribution to the path
integral comes from the "relativistic" quasiparticles. If in addition these
fermions or bosons are  "massless", there  is an extra symmetry: with respect
to the multiplication of the metric by an arbitrary function
$g_{\mu\nu}\rightarrow a^2g_{\mu\nu}$. Such conformal symmetry of
quasiparticles should lead to the conformal invariant and covariant action
for the superfluid condensate, which contains the Weyl tensor (see e.g.
\cite{BirellDavies}).

In a real condensed matter system the covariant terms in the superfluid
action mainly represent the corrections to the leading hydrodynamic terms (see
e.g. \cite{Ilinski}). However for some specific phenomena, for which the
contribution of the low-energy "relativistic" tail is dominating, the
covariant and even conformal terms can be important: anomaly is one of the
examples. In condensed matter the energy cut-off parameter, at which the
low-energy "relativistic" behavior transforms to
the high-energy Galilean one, is the physical parameter and reflects the
fact that the curved space in the low-energy edge is the effective space,
while the underlying coordinate space is pure Galilean. In this flat space
the thermodynamic temperature $T$ is well defined. It is constant in the
global equilibrium and is viewed by the "relativistic" low-energy
quasiparticles as the constant in the Tolman' law for the local
"relativistic" temperature: $T_{\rm eff}({\bf r})\sqrt{g_{00}({\bf
r})}=const=T$. The fact that it is the temperature $T$, which is of the
physical significance, rather than $T_{\rm eff}$, makes physical the
divergence of the quasparticle energy density at horizon. As a result, in
the presence of the event horizon the global thermodynamic equilibrium of
the condensed matter system is impossible: one always has the entropy
production when the normal component of the liquid moves across the horizon.

I am grateful to Dmitrij Fursaev, Ted Jacobson, Alexei Morozov and Alexei
Starobinsky for discussions.

\end{document}